\documentclass{article} % For LaTeX2e
\usepackage{iclr2020_conference,times}

\usepackage{amsmath,amsfonts,bm}

\def\eqref#1{equation~\ref{#1}}
\def\1{\bm{1}}

\DeclareMathAlphabet{\mathsfit}{\encodingdefault}{\sfdefault}{m}{sl}
\SetMathAlphabet{\mathsfit}{bold}{\encodingdefault}{\sfdefault}{bx}{n}

\usepackage{hyperref}
\usepackage{url}
\usepackage{todonotes}

\title{Concrete Problems in AI Safety, Revisited}

\author{Inioluwa Deborah Raji \& Roel Dobbe \\
AI Now Institute\\
New York University\\
New York City, NY, USA \\
\texttt{\{deb,roel\}@ainowinstitute.org} \\
}

\iclrfinalcopy % Uncomment for camera-ready version, but NOT for submission.
\begin{document}

\maketitle

\begin{abstract}
As AI systems proliferate in society, the AI community is increasingly preoccupied with the concept of AI Safety, namely the prevention of failures due to accidents that arise from an unanticipated departure of a system's behavior from designer intent in AI deployment. We demonstrate through an analysis of real world cases of such incidents that although current vocabulary captures a range of the encountered issues of AI deployment, an expanded socio-technical framing will be required for a more complete understanding of how AI systems and implemented safety mechanisms fail and succeed in real life. 

%reveal that accidents are in fact the result of not just technological causes but also heavily influenced by organizational, social and political dynamics outside of the technologist’s control. W

%a grounding in several non-technological causes, and
%As AI systems proliferate society, the AI community is increasingly occupied with the concept of AI Safety, namely the prevention of engineering failures due to accidents that arise in AI deployments. These accidents have been repeatedly defined as a “unintended and harmful behavior that may emerge from poor design of real-world AI systems” - however, an analysis of real world cases reveal that the same accidents may arise from non-technological causes, and in fact, often are the result of organizational, social and political dynamics outside of the technologist’s control. We thus demonstrate that although the vocabulary of this community captures a range of the encountered issues centred on malicious or unanticipated threats to control, an expanded socio-technical framing will be required to capture a more complete understanding of how AI systems and implemented safety mechanisms fail in real life.  
\end{abstract}

\section{Introduction}

The rapid adoption and widespread experimentation and deployment of AI systems has triggered a variety of failures. Some are catastrophic and visible such as in the case of fatal crashes involving autonomous vehicles. Other failures are much more subtle and pernicious, such as the development of new forms of addiction to personalized content. As a response to these safety concerns, a variety of research communities from control theory and formal methods~\citep{fisac_general_2018,seshia_towards_2016}, computer security~\citep{carlini_evaluating_2019,papernot_sok_2018}, engineering for safety assurances~\citep{fang_three_2019,trapp_towards_2019}, and within the AI field itself ~\citep{amodei_concrete_2016,leike_ai_2017,varshney_engineering_2016} have begun to weigh in on how to characterize and address safety risks and vulnerabilities in AI systems. Despite the complexity of modern AI systems and their already widespread impact, few studies are yet to ground their analysis of the taxonomy of problems in an understanding of real world deployments and disasters. 

The goal of this study is to revisit the effectiveness of AI safety problem formulations in providing adequate vocabulary for characterising the key challenges of implementing safe AI systems in the real world. This workshop paper is an initial investigation into ongoing work on this topic.

\section{Methodology}
In this paper, we go through a commonly cited taxonomy for AI safety by \citet{amodei_concrete_2016}, who survey and taxonimize a body of 170 papers to propose a subset of \emph{concrete AI Safety problems} to pay attention to, aiming to understand ``what challenges remain to reducing accident risk in modern machine learning systems''. In this context, an accident is defined as ``a situation where a human designer had in mind a certain (perhaps informally specified) objective or task, but the system that was designed and deployed for that task produced harmful and unexpected results.'' The authors then propose to ``categorize safety problems according to where in the [technical AI design] process things went wrong.''  

%Recent work has analyzed how the ML and AI communities tend to \emph{internalize} issues such as safety or other value conflicts within the process and technical elements of a system rather than framing these in their broader sociotechnical context~\citep{dobbe_hard_2019,green_algorithmic_2020}. 

%In this position paper we support these claims, by showing that, while many accidents can be \emph{interpreted} as a concrete safety problem, evidence for their root causes shows that \emph{resolutions require interventions that extend beyond any technical solution} affiliated with the concrete problem categories.

We re-examine the provided definitions of three concrete problems from the taxonomy presented by Amodei et al., and identify a real world use case that fits the provided descriptions of the safety risk. We then diagnose the nature of the challenges in the real world, paying particular attention to dimensions of the issue that current frameworks do not adequately address. 
%\todo[inline]{you mean do not?}

%Case studies are selected due to their sufficiently detailed coverage and reporting - in incident reports, lawsuits, investigative journalism features and research papers -  with an articulated harm and articulated probable cause following some form of credible public investigation. Open cases and ongoing disputes, or cases in which the impact is uncertain will not be considered.  

%The goal of this study is to revisit often cited framings of AI safety and understand how effective such problem formulations are in terms of providing adequate vocabulary for characterising the key challenges of implementing safe AI systems in the real world and minimizing the harm caused by currently deployed AI systems. 

%\subsection*{Avoiding Negative Side Effects}
\subsection*{Safe Exploration}

\defcitealias{uber-crash}{(NTSB~\citeyear{uber-crash})}%
\defcitealias{uber-crash-2}{(NTSB~\citeyear{uber-crash-2})}%

\defcitealias{tesla-crash}{(NTSB~\citeyear{tesla-crash})}%

\defcitealias{tesla-crash-2}{(NTSB~\citeyear{tesla-crash-2})}%

%\textbf{\textit{Safe Exploration:}} 
Safe exploration is the minimization of undesirable behaviour or harm arising from the learning process of an algorithm. For instance, during the training process of a machine learning model or while accumulating training examples or exploring a reinforcement learning environment ~\citep{amodei_concrete_2016,pecka2014safe}. Autonomous exploration is often complicated by the unpredictable nature of human behavior, which may amount to a prematurely deployed system with no sensible safety guarantees. Exploration risks are thus on par with that of untested systems released in the wild - and the justification for taking such risk is itself an ethical dilemma~\citep{bird2016exploring}.

A clear example of this phenomenon can be observed with the development of autonomous vehicles. Self-driving vehicle companies boast in safety reports and in marketing material about the use of data from a fleet of deployed vehicles to fine-tune their algorithms. Some providers, such as Tesla, harness the power of their client vehicles, taking data from sold cars to train new models and then release synchronous updates for improved navigation in all vehicles ~\citep{teslasafe}. 
Others, in a more experimental phase of development, such as Uber, hire test drivers to log thousands of miles, collecting data for developing models and designing new algorithms for future deployment in sold vehicles ~\citep{ubersafe}. In both cases, the safety risk of such explorations is clear - testing a car in the public sphere is inherently dangerous, risking the lives of pedestrians and fellow drivers in order to collect the data used to develop navigation systems. While on a test drive, an Uber autonomous vehicle killed a woman in Tempe, Arizona, on March 18, 2018 ~\citetalias{uber-crash}. Similarly, the autopilot feature of the Tesla Model S ~\citetalias{tesla-crash} caused a fatal crash on May 7, 2016, near Williston, Florida. Public facing journalistic reporting on these incidents focused on the particularities of the engineering failures that led to these fatal crashes - Uber’s algorithm did not identify the pedestrian crossing outside a crosswalk, and Tesla’s system did not distinguish between the sky and the white side of the truck it crashed into. However, these products were effectively in public beta-testing. They were \emph{expected} to fail. %Of true scrutiny should be the effectiveness of the \emph{safety measures} employed, such as fail-safe features to kick in when an expected engineering failure occurs. Inadequate safety measures were what contributed most to these accidents and 
It was thus not these failures directly, but the lack of adequate \emph{safety measures} that was ultimately the focus of what the National Transportation Safety Board (NTSB) declared as the probable cause of the accidents in both cases. 

At first glance, the case of Uber ressembles the scenario of ``absent supervision''~\citep{leike2017ai}. As the operator was visually distracted throughout the trip by a personal cell phone, it seems as though the primary issue could be a lack of contingency planning for situations where the driver was ``absent'', ie. not in a position to intervene or react as expected to correct for a technical failure. However, in reality, the NTSB found that Uber ATG did have such an oversight functionality in place. Managers “had the ability to retroactively monitor the behavior of vehicle operators", and notify them when they perceived a need to react. However, ``they rarely did so'', leading the NTSB to declare the official cause of the accident to be the company’s “inadequate safety culture”, which normalized the casual treatment and ultimate neglect of much of the nominally established safety protocol ~\citetalias{uber-crash-2}. 

The Tesla case reveals a similar situation. According to the NTSB report, at the time of the crash, the vehicle could not “recognize the impending crash. Therefore, the system did not slow the car, the forward collision warning system did not provide an alert, and the automatic emergency braking did not activate” ~\citetalias{tesla-crash-2}.
Thus, the car would not stop until the driver overrode the system to stop - a safety mechanism understood as “safe interruptibility” ~\citep{orseau2016safely}. 
However, as NTSB reported, %"the way in which the Tesla 'Autopilot' system monitored and responded to the driver’s interaction with the steering wheel was not an effective method of ensuring driver engagement"~\citetalias{tesla-crash-2}.
despite the driver's ability to interrupt the system effectively, once within range of fatal risk, the bewildered and terrified driver was not prepared to take any action at all. "Overreliance on Automation" was thus named amongst the key causes for the accident~\citetalias{tesla-crash-2} as "the way in which the Tesla 'Autopilot' system monitored and responded to the driver’s interaction with the steering wheel was not an effective method of ensuring driver engagement"~\citetalias{tesla-crash-2}. This is a well-studied phenomenon in autopilots and human-machine interaction~\citep{parasuraman_humans_1997} - when the limits of the system are not clearly communicated, it can become difficult for the human to understand when they are more qualified than the algorithm to make a decision, and remain alert to any changes to their role in decision making.  “Tesla driver’s pattern of use of the Autopilot system indicated an over-reliance on the automation and a lack of understanding of the system limitations”~\citetalias{tesla-crash}. 
%In response to the NTSB report, Tesla redesigned its “Autopilot” system, reducing the "period of time before the Autopilot system issues a warning/alert when the driver’s hands are off the steering wheel" and adjusting alert timing to ensure drivers remained engaged. %in the case of  %recommendation was thus for Tesla to make  following this re

%The Tesla driver’s pattern of use of the Autopilot system indicated an over-reliance on the automation and a lack of understanding of the system limitations.

%Tesla made design changes to its “Autopilot” system following the crash. The change reduced the period of time before the “Autopilot” system issues a warning/alert when the driver’s hands are off the steering wheel.  The change also added a preferred road constraint to the alert timing sequence

%\todo[inline]{Use reference: Over reliant on automation:
%\url{https://www.sfchronicle.com/business/article/Tesla-Autopilot-let-driver-rely-too-much-on-14414600.php} ``It’s unrealistic to try to train people for automation,'' Friedman said. ``You’ve got to train automation for people.''
%}
In both cases, it was thus a lack of stakeholder engagement with the safety feature, rather than the technical function of the safety feature itself that were deemed primary causes of the accidents.

\subsection*{Avoiding Negative Side Effects}

%\textbf{\textit{Avoiding Negative Side Effects:}} 
At times, an AI system can cause unintentional and unknown side effects. These are often described as phenomena that are unanticipated indirect results of the algorithmic system’s deployment, most specifically the unintended consequences of an agent’s actions within a specific environment~\citep{amodei_concrete_2016}. 

Sometimes the reward beneficial to the actor creating and deploying the algorithm is inherently harmful to another population. %and, given their power to control the algorithm, developers can prioritize their own benefit to the detriment of other actors. 
This is complementary to the view of side effects as ``externalities'' inherently detrimental to other stakeholders within what is truly a multi-agent system ~\citep{overdorf2018pots, amodei_concrete_2016}. However, reality reveals the inherent power imbalance between modeled agents. Those with control of the system - and likely greater social and financial capital - can thus dictate the values and priorities of the systems they choose to build, at times in knowing direct opposition to the well-being of other, less influential actors. For instance, a class action lawsuit “Doe v. Netflix” reveals the tension between an exposure of private personal records and the improved accuracy of the platform's recommendation engine through the Netflix Prize ~\citep{singel2009netflix}. As video records are actually amongst the most protected personal records in the United States, there was legal recourse to make such a case against corporate interests to protect the potential harm to subscribers ~\citep{singel2010netflix}. Without such legal responsibility, however, it is unclear if, given the advances in accuracy and product quality ~\citep{gomez2015netflix}, the company would have been as willing to compromise, despite the clear threats to consumer privacy ~\citep{calandrino2011you,narayanan2006break}. For instance, other platform issues involving less legal repercussion - such as screen addiction, especially amongst impressionable youth ~\citep{matrix2014netflix} - have yet to meaningfully influence corporate decision-making.

Additionally, unlike the focus of the current framing, it is not just the \emph{actions} of an AI agent that can produce side effects. In real life, basic design choices involved in model creation and deployment processes also have consequences that reach far beyond the impact that a single model’s decision can have. In reality, for AI systems to even be built, there is very often a hidden human cost ~\citep{gray2019ghost, birhane2020robot}. These \emph{harms of production} are not emergent harms from the training process, as we see with safe exploration, but rather a byproduct of the resource requirements for the creation of a model. Data requirements, compute requirements, API model delivery decisions can all invite certain harms to privacy, sustainability and corporate accountability. %Architectural and infrastructural design decisions generate side effects. 
For example, deep learning approaches to facial recognition requires by design widespread privacy violations ~\citep{raji2020saving}, as millions of faces - which under ISO standards are considered sensitive identifiable biometric information - are required for satisfactory training and testing of the models. Similarly, poorly paid and trained annotators are often exploited for the sake of doing the mundane work of labeling large amounts of at times sensitive and graphic data %, including traumatic graphic images for moderation models and intimate domestic conversations
~\citep{gray2019ghost}. Our training and testing environments for models intended for real world deployment is often the byproduct of human exploitation - whether through direct labour or by harvesting our data footprints. Similarly, such large scale data storage and compute costs from the current trend of AI development invites climate abuse ~\citep{NLPeng, greenai, showwork}, a natural crisis also disproportionately impacting the often lower income and thus most neglected humans in society. 

There is thus a need to consider the power dynamics at play in the development and deployment of these systems and begin to define the value tradeoffs that ultimately determine a system’s outcomes and broader implications, beyond the context of an assumption of benevolent designer intent. %towards more decentralized and democratic models of control and contribution, as well as 

\subsection*{Scalable Oversight}
%\textbf{\textit{Scalable Oversight:}}
Scalable oversight refers to situations in which a safety risk is so infrequent, or the objective function so nebulous, that it becomes too expensive to evaluate frequently~\citep{amodei_concrete_2016}. 
%IEEE report:https://spectrum.ieee.org/biomedical/diagnostics/how-ibm-watson-overpromised-and-underdelivered-on-ai-health-care
%
%Nevertheless, the consequences are large and the cost of failure requires a serious consideration of 
%
%TO DO (Deb) - talk about failure 
%In 2012, the University of Texas M. D. Anderson Cancer Center in Houston partnered with IBM to develop the artificial intelligence program, called IBM Watson, as a clinical decision tool in oncology. Five years and $62 million later, M. D. Anderson let its contract with IBM expire before anyone used Watson on actual patients.
%
In AI for healthcare, a heavily regulated safety-critical domain, the real world context is often hard to access. This motivates technologists to design and train the system using proxy or simulated objectives that may overlook or fail to adequately represent contextual factors. For instance, an internal investigation by the University of Texas M. D. Anderson Cancer Center in Houston found the IBM Watson’s Oncology Expert Advisor tool, into which they had invested  \$62 million over 5 years, to fall short of the “gold standard” of medicine ~\citep{report-healthcare}. Unable to access patient records due to health privacy laws, Watson diagnosis and treatment recommendations are informed from official medical guidelines and peer-reviewed studies from academic medical journals - rather than patient records of experience. 

%Additionally, machine learning as a method does not effectively accomodate outlier inputs. %Even when the outlier case does arrive and the human is given flexibility to taking control, automation bias occurs ~\citep{mosier1998automation}. In handling outliers, humans tend to want to trust the robot over themselves even in situations when the human is the more reliable expert and meant to override an algorithm's prediction in order to address issues ~\citep{dobbe2018broader}. 

As a result, not only was the system often too general to be helpful in clinical practice, but the system was unable to adapt to new clinical situations in a way that put patients at risk of lower quality care. For instance, the 2018 breakthrough discovery of a dramatically effective new “tissue agnostic” cancer drug for cancer tumors containing a specific genetic mutation led to the drug’s release being fast tracked. It was subsequently approved by the FDA with only results available in 55 patients, 4 of whom had lung cancer. Following the finding, all prior guidelines for lung cancer treatments were revised, now with an emphasis on testing for the genetic mutation that may qualify patients for the novel treatment. However, by virtue of the nature of its machine learning system, IBM Watson would not change its predictions based on these four patient cases~\citep{strickland2019ibm}.
As a result, the program shut down without any physicians having the confidence to use the tool on actual patients ~\citep{schmidt2017md}.  %- the advice provided through . Additionally, .  It was subsequently approved by the FDA with only results available in 55 patients, 4 of whom had lung cancer. Following the finding, all prior guidelines for lung cancer treatments were revised, now with an emphasis on testing for the genetic mutation that may qualify patients for the novel treatment. However, IBM Watson would not change its predictions based on just four patient cases~\citep{strickland2019ibm}. %The lack of the integration of contextual though unstructured patient case data and the tool's inability to adapt to new clinical situations ultimately made it too challenging to assimilate Watson into the hospital setting. 

Resorting to proxies or synthetic data comes with its own assumptions and risks which should be carefully accounted for in safety-critical environments. Alternatively, one could acknowledge that certain use cases are more suitable to the method of machine learning and others are more unsafe and volatile when addressed using this method. Within IBM Watson for Healthcare, the IBM for Genomics project provides such an example ~\citep{strickland2019ibm}. Genetic information is structured and consistently recorded yet unique to each patient - the goal of needing an aid for pattern recognition on large amounts of structured data fits the paradigms and assumptions under which ML systems are trained and tested. As a result, this has been the most successful Watson launch to date. %and is effective in supporting clinicians with genetic classification and testing for patient treatment design. 

\section{Recurring Themes in Real World Safety Failures}

% 1. There is a Difference Between Flaws in Theoretical Design and Errors in Engineering Practice %The implementation and maintenance of a system will introduce safety challenges, which are often ignored - any theoretical system will still need to be engineered, now and in the future and the challenges of engineering often introduce safety problems 

% 2. Theory is grounded in speculative or deductive reasoning rather than being regularly validated through inductive reasoning. Equations representing the wrong problem, variables that engineers are not actually in control of. 

%3. Looking at broader stakeholders impact and not just designer intent to shape the motives of the problem. Need to adopt a socio-technical framing of what safety and vulnerabilities are. Neglecting realities of power structures and stakeholder negotiation, meaningful contestability as a means of ensuring what's best for even non-users, as well as users.

Throughout the various case studies and analyses in the previous section, a set of themes surfaced. %about which causes and mechanisms proved to be critical in the occurrence of the failure or essential in improving the situation. 
These themes are offered for discussion to inspire future work.

\textbf{(1) We must consider errors in engineering practice, not just 
flaws in theoretical design or formalization.} Any theoretical AI system will need to be implemented in some way to have an impact on the real world. At times, it is the mistakes made in the process of its implementation that may lead to an accident. The reality of the engineering processes - including task design, specification, construction, validation, integration and maintenance - that arise in practice are often neglected in current explorations of AI safety, even though this is when safety problems are often most pronounced and best addressed. %Current explorations of AI safety 
%Without verifying and validating a system in practice, it is hard, if not impossible, to provide adequate theoretical foundation. For safety, practice and theory should strengthen each other.

\textbf{(2) We need to validate safety problems through inductive reasoning.} A theoretical or speculative model can only go so far in representing a problem adequately. Properly safeguarding a system requires mechanisms that are developed inductively through ongoing validation within the context of real world case studies. Otherwise, failures will arise from the misrepresentation of the extent and nature of engineering decision-making or the influence of stakeholder participation. %Equations representing the wrong problem, variables that engineers are not actually in control of. 

\textbf{(3) We must consider broader stakeholder impact and interactions, not just alignment with technical design intent when considering accidents.} AI systems mediating sensitive domains and public spaces require broad deliberation and corroboration of safety requirements across different stakeholders, rather then being decided by a small group of developers and entrepreneurs. This requires a socio-technical framing that is cognizant of power structures, and provides meaningful forms of dissent across affected communities.

\section{Conclusion}

These case studies demonstrate that the failure of an AI system in the real world can differ significantly from the expectations set by our own taxonomies of what it means for an AI system to be safe. All too often the focus of AI safety is on the developer's control - over the reward function, the training circumstances and the agent's impact in a given environment. However, in the real world, failures take on a new dimension of complexity and reveal themselves to be inherently systematic rather than contained within any technological artifact. The reported cause of many of these cases are hardly ever attributed to a technological malfunction but rather a network of ineffective socio-technical interactions with users and other stakeholders.

\bibliographystyle{iclr2020_conference}
%\bibliographystyle{plain} 
%\bibliography{refs_concreteproblems}
\bibliography{iclr2020_conference}

% \appendix

\end{document}